\documentclass[sigconf, pbalance]{acmart}

\usepackage{multirow}
\usepackage{graphicx}
\usepackage{booktabs}
\usepackage{multirow}
\usepackage{subfigure}
\usepackage{wrapfig}
\usepackage{pifont}
\usepackage{soul} 
\usepackage{amsmath,amsfonts,bm}

\def\eqref#1{equation~\ref{#1}}

\def\1{\bm{1}}

\DeclareMathAlphabet{\mathsfit}{\encodingdefault}{\sfdefault}{m}{sl}
\SetMathAlphabet{\mathsfit}{bold}{\encodingdefault}{\sfdefault}{bx}{n}

\usepackage{hyperref}
\usepackage{url}
\usepackage{xspace}

\acmSubmissionID{922}

\newcommand{\name}{\textsc{DiffSound}\xspace}

\definecolor{nicegreen}{rgb}{0.1, 0.6, 0.2}

\title[\name: Differentiable Modal Sound Rendering and Inverse Rendering]{\name: Differentiable Modal Sound Rendering and Inverse Rendering for Diverse Inference Tasks}

\author{Xutong Jin}
\authornote{Xutong Jin and Chenxi Xu have equal contributions.}
\email{jinxutong@pku.edu.cn}
\affiliation{%
  \institution{School of Computer Science, Peking University}
 \country{China}
}
\author{Chenxi Xu}
\authornotemark[1]
\email{2301213239@pku.edu.cn}
\affiliation{%
  \institution{School of Computer Science, Peking University}
 \country{China}
}
\author{Ruohan Gao}
\email{rhgao@umd.edu}
\affiliation{%
  \institution{University of Maryland, College Park}
 \country{U.S.A}
}
\author{Jiajun Wu}
\email{jiajunwu@cs.stanford.edu}
\affiliation{%
  \institution{Stanford University}
 \country{U.S.A}
}
\author{Guoping Wang}
 \email{wgp@pku.edu.cn}
\affiliation{%
  \institution{School of Computer Science, Peking University}
  \country{China}
}

\author{Sheng Li}
\authornote{Sheng Li is the corresponding author.}
\email{lisheng@pku.edu.cn}
\affiliation{%
  \institution{School of Computer Science, Peking University}
  \country{China}
}

\copyrightyear{2024}
\acmYear{2024}
\setcopyright{acmlicensed}\acmConference[SIGGRAPH Conference Papers '24]{Special Interest Group on Computer Graphics and Interactive Techniques Conference Conference Papers '24}{July 27-August 1, 2024}{Denver, CO, USA}
\acmBooktitle{Special Interest Group on Computer Graphics and Interactive Techniques Conference Conference Papers '24 (SIGGRAPH Conference Papers '24), July 27-August 1, 2024, Denver, CO, USA}
\acmDOI{10.1145/3641519.3657493}
\acmISBN{979-8-4007-0525-0/24/07}


\begin{document}

\begin{abstract}
Accurately estimating and simulating the physical properties of objects from real-world sound recordings is of great practical importance in the fields of vision, graphics, and robotics. However, the progress in these directions has been limited---prior differentiable rigid or soft body simulation techniques cannot be directly applied to modal sound synthesis due to the high sampling rate of audio, while previous audio synthesizers often do not fully model the accurate physical properties of the sounding objects. We propose \name, a differentiable sound rendering framework for physics-based modal sound synthesis, which is based on an implicit shape representation, a new high-order finite element analysis module, and a differentiable audio synthesizer.
Our framework can solve a wide range of inverse problems thanks to the differentiability of the entire pipeline, including physical parameter estimation, geometric shape reasoning, and impact position prediction.
Experimental results demonstrate the effectiveness of our approach, highlighting its ability to accurately reproduce the target sound in a physics-based manner. \name serves as a valuable tool for various sound synthesis and analysis applications.
\end{abstract}


\begin{teaserfigure}
\centering
 \includegraphics[trim={0cm, 0cm, 0cm, 0cm},clip,width=0.9\linewidth]{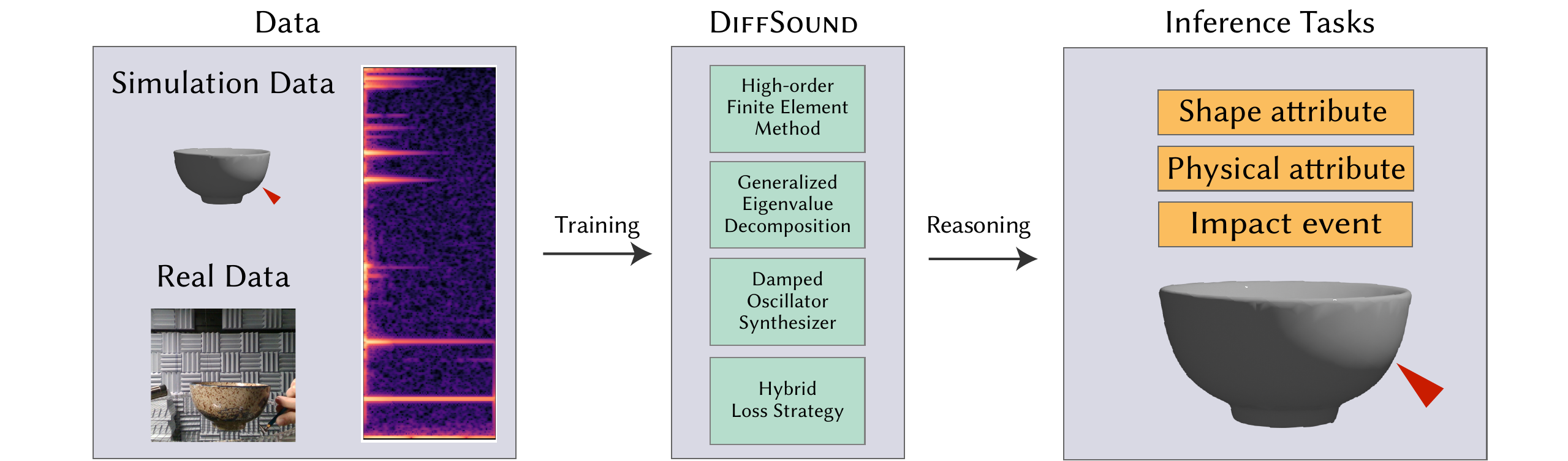}
\caption{
We introduce \name, a differentiable sound rendering framework for physics-based modal sound synthesis. It can infer a variety of physical and shape attributes, such as material, volumetric thickness, geometric shape, and impact positions of the object, from both simulated datasets and real sound recordings, enabling a series of inverse rendering applications. 
}
\label{fig:teaser}
\end{teaserfigure}

\maketitle

\section{Introduction}

The concept of differentiable simulation has become increasingly popular in the graphics and machine learning communities in recent years~\citep{Popovi2003, Avila2018, Toussaint2019, degrave2019, Qiao2020, Jie2021}. A differentiable simulation framework has the benefit of allowing for gradient-based optimization and thus can be easily integrated into a neural network for end-to-end learning.

Our work focuses on differentiable sound rendering, which addresses a unique challenge compared to standard differentiable rigid or soft body simulations~\citep{hu2020, Geilinger2020, Du2022, degrave2019, Qiao2020, Jie2021} due to the high sampling rate of sound. While previous audio synthesizers~\citep{engel2020ddsp, clarke2021diffimpact} can optimize for many audio and physical-based properties, they are unable to explicitly model more fundamental physical properties such as Young's modulus, Poisson's ratio, size, or shape of the object, and the impact position, which are all critical for realistic modal sound synthesis.


Inferring these objects' properties from real sound recordings can potentially enable various Real-to-Sim applications. For example, we can accurately infer material parameters from real-world recordings and use them to re-create realistic virtual objects, such as those in \citet{gao2021objectfolder,gao2022ObjectFolderV2,gao2023ObjectFolderBM,clarke2023realimpact}. We can also leverage a differentiable sound rendering framework to design the shape and material of virtual objects to produce the desired sound, and then transfer the results back to real objects using 3D printing technology~\citep{Gaurav2015}. The information about an object's shape, material, and impact position can also complement visual perception, particularly in cases of low visual resolution or poor lighting, especially for multisensory robotic applications~\citep{clarke2021diffimpact,li2022seehearfeel}.

Towards this end, we introduce \name, a differentiable simulation framework for modal sound synthesis that employs a high-order finite element method for physics-based modeling. It not only is fully-differentiable and allows for efficient end-to-end optimization, but also establishes a seamless connection between the recorded audio and the fundamental physical properties of real-world objects.
Our \name consists of three main components. First, we propose a hybrid shape representation that combines implicit neural representation and explicit 3D tetrahedral mesh representation. Second, we introduce a high-order finite element analysis module that allows for incorporating different material and shape parameters. Finally, we design a differentiable audio synthesizer with a hybrid loss strategy to enable smooth optimization of the entire differentiable simulation pipeline.

We demonstrate the effectiveness of our method through a wide range of inverse rendering tasks. Our differentiable framework can accurately estimate the attributes of the sounding objects, such as the material parameters, identify the impact positions and amplitudes of the physical interactions, and infer the object shape characteristics, including volumetric thickness, and geometric form, on both synthetic and real-world audio datasets. Notably, while limited prior work has proposed to infer material properties from sound~\cite{Ren2013}, to our best knowledge, our work marks the first attempt to estimate an object's thickness, precise geometric shape and impact position purely through sound analysis. 


\section{Related work}


\paragraph{Modal Sound Synthesis} 
Modal sound synthesis is a technique that has been used to synthesize sounds of rigid bodies~\citep{van2001, OBrien2002, Raghuvanshi2006}. These methods compute the vibration modes of a 3D object through a generalized eigenvalue decomposition. Based on the basic modal sound method, many complex sound phenomena can be simulated, such as knocking, sliding, and friction sound~\citep{van2001}, acceleration noise~\citep{Chadwick2012}, complex damping sound~\citep{Sterling2019}, and high-quality contact sound~\citep{Zheng2011}.

Related to prior studies on estimating material parameters from pre-recorded audio \citep{Ren2013, Zhang2017}, our work differs by providing an end-to-end optimization-based solution, enhancing accuracy. Unlike previous methods optimizing object shape for desired sound \citep{Gaurav2015}, our approach optimizes all sound modes, not just the fundamental frequency. Additionally, it offers greater flexibility in shape optimization, surpassing simple scaling and stretching.

\paragraph{High-Order FEM}
In engineering, higher-order methods are often preferred over lower-order methods due to their superior accuracy and convergence properties. In computer graphics, finite element methods (FEM) with linear shape functions are prevalent due to their simplicity and computational efficiency. While limited prior work demonstrates that higher-order methods have the potential to produce better simulation results~\citep{Mezger2008, Bargteil2014, schneider2019poly, Longva2020}, they are not commonly used in the field.

To the best of our knowledge, the sole previous attempt~\citep{Gaurav2015} that incorporates high-order FEM in modal sound synthesis directly employs the engineering software COMSOL~\citep{comsol} to obtain the results. 
In contrast, within our differentiable framework, we newly implement a high-order FEM module to guarantee both high-quality sound rendering and differentiability.

\paragraph{Differentiable Simulation}
Differentiable simulation has recently gained much popularity in the graphics and machine learning communities. Several advances have been made in this field with differentiable simulators designed for rigid-body dynamics~\citep{Popovi2003, Avila2018, Toussaint2019, degrave2019, Qiao2020, Jie2021, simon2023ral}, soft-body dynamics~\citep{Hu2019, Hahn2019, hu2020, Geilinger2020, Du2022}, fluid dynamics~\citep{Treuille2003, McNamara2004, Wojtan2006, schenck2018, Holl2020}, and cloth~\citep{Liang2019, murthy2021, Li2022}. 

There are also differentiable rendering methods proposed for signal processing~\citep{engel2020ddsp} and modeling impact sound~\citep{clarke2021diffimpact}. These methods can capture various physics-based properties, such as modal response and force profiles. However, they do not explicitly consider the fundamental physical properties of objects, such as shape, material, and impact position. Another promising approach uses neural networks to approximate the modal analysis process~\citep{Jin2020, Jin2022}. Although neural networks are inherently differentiable, ensuring physical accuracy can be challenging, and accurate modal analysis cannot be trivially achieved just through neural network optimization. 

\vspace{0.1in}
\section{Differentiable Modal Sound Rendering}
First, we give an overview of our full model (Sec. \ref{sec:framework_overview}). Then, we describe the differentiable tetrahedral mesh representation (Sec. \ref{sec:diff_tetra_mesh}), differentiable high-order finite element method (FEM) for modal analysis  (Sec. \ref{sec:diff_fem}), and hybrid loss strategy for optimizing all learnable modules (Sec. \ref{sec:loss_func}).

\begin{figure*}[t]
    \centering
    
    \includegraphics[width=0.9\linewidth]{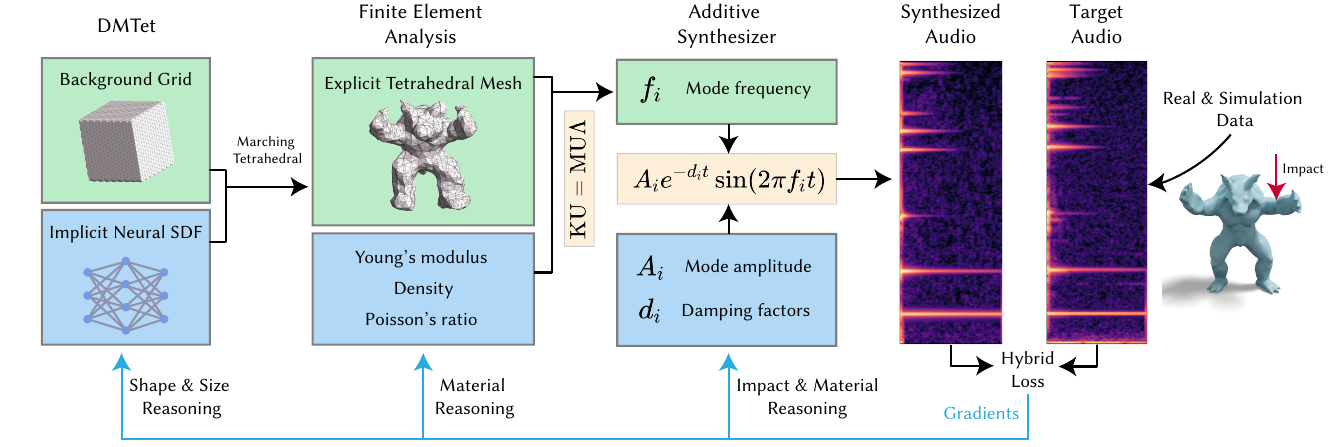}
    \vspace{-0.1in}
    \caption{Our \name differentiable simulation and inverse rendering pipeline. The differentiable tetrahedral mesh representation is employed to directly optimize the topology of a tetrahedral mesh. Subsequently, a differentiable high-order finite element analysis module is utilized to analyze the vibration frequencies of the tetrahedral mesh. Finally, a differentiable additive synthesizer is used to produce the impact sound with a hybrid loss function for optimizing all learnable modules. The learnable parameters, indicated by blue boxes, control module outputs in our differentiable framework. This enables gradient computation for hybrid loss, facilitating parameter optimization. 
 }
 \vspace{-0.1in}
    \label{fig:overall}
\end{figure*}

\subsection{Method Overview}
\label{sec:framework_overview}

Differentiable methods in machine learning, like our sound synthesis method \name, allow for computing output gradients from model parameters. This facilitates parameter optimization using gradient-based algorithms, with our method ensuring a differentiable pipeline from 3D mesh input to modal sound output.

Specifically, assuming $\theta$ represents the learnable parameters in our framework. For an input mesh $m$, we introduce a differentiable explicit tetrahedral mesh generator $G_{\theta_G}$, which transforms the input mesh into an explicit tetrahedral mesh $m_{tet} = G_{\theta_G}(m)$. The generated tetrahedral mesh is constrained by generator parameters $\theta_G$, such as geometric shapes and thickness. 

Next, we introduce a differentiable high-order FEM module, including a differentiable FEM matrix assembler $A_{\theta_A}$ and a differentiable generalized eigenvalue decomposer $D$. The matrix assembler takes an explicit tetrahedral mesh as input, and outputs its mass matrix and stiffness matrix $\mathbf{M}, \mathbf{K} = A_{\theta_A}(m_{tet})$ constrained by parameters $\theta_A$, including Young's modulus and Poisson's ratio. The generalized eigenvalue decomposer $D$ takes $\mathbf{M}, \mathbf{K}$ as input and outputs its eigenvalues corresponding to $\mathbf{K}\mathbf{U} = \mathbf{M}\mathbf{U}\mathbf{\Lambda}$, denoted as $\lambda = diag(\mathbf{\Lambda}) = D(\mathbf{M}, \mathbf{K})$. 

Building upon this, we introduce a differentiable additive synthesizer $S_{\theta_S}$, which takes eigenvalues $\lambda$ as input and synthesizes its modal sound $a_{syn} = S_{\theta_S}(\lambda)$ with the constraint of $\theta_S$, such as damping coefficients and mode amplitudes. 
Finally, we introduce a hybrid loss function $L$ that compares the synthesized audio $a_{syn}$ generated from the aforementioned process with the ground-truth audio $a_{gt}$, resulting $loss = L(a_{syn}, a_{gt})$. For simplicity, we consolidate the process of differentiable sound synthesis by the mesh $m$ into a function $F$: $a_{syn} = F_{\theta}(m) = S_{\theta_S}(D(A_{\theta_A}(G_{\theta_G}$
$(m))))$. At this point, the optimization target is as follows: 
\begin{equation}
\theta^* = \text{argmin}_{\theta} (L(F_{\theta}(m), a_{gt})).
\end{equation}

In certain tasks, it is often required to optimize the parameters of just one module while the parameters of other modules remain fixed. The schematic representation of our approach's methodology is depicted in Figure \ref{fig:overall}.

\subsection{Differentiable Tetrahedral Representation}
\label{sec:diff_tetra_mesh}
We propose a differentiable tetrahedral mesh representation tailored for our differentiable sound rendering, building upon the foundation of Deep Marching Tetrahedra (DMTet)~\citep{shen2021deep,munkberg2022extracting}. Our approach involves the representation of a shape through a Signed Distance Field (SDF) implicitly encoded by a Multilayer Perceptron (MLP) (Sec. \ref{sec:implicit_neural_representation}), which is then transformed into an explicit tetrahedral mesh using a deformable tetrahedral grid (Sec. \ref{sec:implicit_to_explicit}).

\subsubsection{Implicit Neural Representation}

\label{sec:implicit_neural_representation}

Given the inherent challenge of precisely associating the sound of an object with its exact shape, there can be significant ambiguity in the resulting geometry when optimizing purely from sound. To address this, we utilize a Multilayer Perceptron (MLP) to parameterize the SDF values. This implicit parameterization effectively serves to regulate both the SDF and the overall smoothness of the reconstructed shape. Additionally, the degree of smoothness can be fine-tuned by varying the frequency of the positional encoding, following Neural Radiance Fields~\citep{mildenhall2020nerf}, applied to the inputs of the MLP.

\begin{figure}[th]
    \centering
    \includegraphics[width=1.0\linewidth]{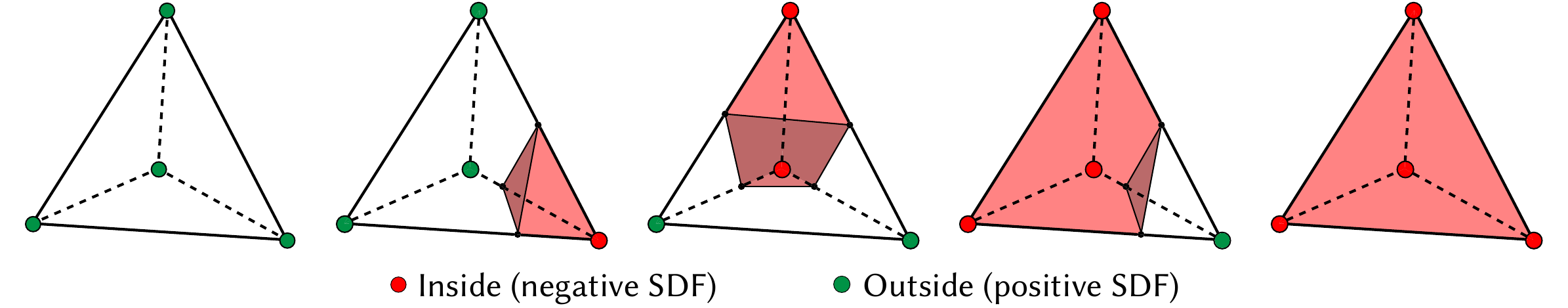}
    \caption{Five configurations of the interface between background tetrahedrons and internal ones. If the internal subregion is more complex than a tetrahedron, it should be subdivided into smaller tetrahedrons.}
    \label{fig:dmtet}
\end{figure}

\subsubsection{Implicit to Explicit Representation}
\label{sec:implicit_to_explicit}

We adopt the Marching Tetrahedra (MT) algorithm \citep{marchingtet} to transform encoded Signed Distance Function (SDF) data into explicit tetrahedral meshes. Our approach allows background tetrahedral cell vertices to deform within half-cell size limits, enhancing geometric expression. Using SDF values from the MLP for vertices in a tetrahedron, MT discerns surface topology based on SDF sign variations. Our method focuses on identifying internal tetrahedra rather than surface topology, resulting in five configurations due to rotational symmetry (see Figure \ref{fig:dmtet}). Surface vertex locations are determined by linear interpolation along tetrahedron edges, akin to DMTet's approach \citep{shen2021deep,munkberg2022extracting}. For complex internal sub-regions, we further subdivide into smaller tetrahedrons. To minimize high-frequency noise impact in sound optimization, we extract the largest connected tetrahedral mesh, discarding small fragments. Despite the potential sudden appearance or disappearance of fragments, their existence is consistently determined during the computation of the gradient in each step. Consequently, our approach ensures the accurate calculation of the gradient.


\subsection{Differentiable High-order FEM}
\label{sec:diff_fem}
Prior studies~\citep{hughes2012, Gaurav2015} have noted the limitations of linear tetrahedral finite elements in producing accurate solutions, even with refined simulation discretization. In this work, we propose the use of differentiable high-order FEM for greater accuracy and generality.

We compute the mass and stiffness matrices for the tetrahedral mesh (introduced in Sec. \ref{sec:diff_tetra_mesh} above), which are made differentiable with respect to the material coefficients, namely Young's modulus, density, and Poisson's ratio as introduced in Sec. \ref{sec:mass_stiffness_matrix}. Subsequently, in Sec. \ref{sec:eigenvalue_decomp}, we compute the gradient from the eigenvalues obtained through eigendecomposition with respect to these two matrices.
For a comprehensive derivation of these matrices, please refer to~\citep{FEMcourse, zhu2018finite}.

\subsubsection{Mass and Stiffness Matrix}
\label{sec:mass_stiffness_matrix}
To obtain the mass matrix, we initially compute the element matrix for each individual tetrahedral element, followed by the assembly process to construct the mass matrix for the entire tetrahedral mesh.
Let $V$ denote the volume occupied by a tetrahedral element, $\rho$ represents its density, and the shape function value at position $x$ with respect to node $i$ is denoted as $N_i(x)$. The element mass matrix $\mathbf{M}_e$ is defined as follows:
\begin{equation}
\mathbf{M}_e^{ij} = \rho \iiint_{x \in V} N_i(x)N_j(x)dx \ .
\label{eq:mass_matrix_element}
\end{equation}
To compute this volume integral, we employ the Gaussian numerical integration method, selecting $t$ Gaussian integration points $g_k$ within the tetrahedral element, with corresponding Gaussian integration weights $w_k$. The unit mass matrix can be calculated as:
\begin{equation}
\mathbf{M}_e^{ij} = \rho V \sum_{k=1}^t N_i(g_k)N_j(g_k)w_k \ .
\label{eq:mass_matrix_unit}
\end{equation}
For a high-order tetrahedral element containing $n$ nodes, the algorithm described above yields a unit mass matrix $\mathbf{M}_e$ of size $3n \times 3n$. Now, for the entire tetrahedral mesh with a total of $m$ nodes, it is only necessary to add each element $\mathbf{M}_e^{ij}$ computed for each tetrahedron to the corresponding entries $\mathbf{M}^{ij}$ of the overall mesh's mass matrix $\mathbf{M}$. This assembles a $3m \times 3m$ mass matrix $\mathbf{M}$.

Following the defined process for the mass matrix, let $E$ denote Young’s modulus and $\nu$ denote Poisson's ratio. The element stiffness matrix $\mathbf{K}_e$ of size $3n \times 3n$ for a tetrahedral element is defined as:
\begin{equation}
\mathbf{K}_e = \sum_{k = 0}^t w_k V \mathbf{D}(g_k)^T \mathbf{B}(E, \nu) \mathbf{D}(g_k) \ .
\label{eq:element_stiffness}
\end{equation}
Here, $\mathbf{B}(E, \nu)$ is the elasticity matrix representing the material model, and we adopt the linear elastic model~\citep{FEMcourse}. $\mathbf{D}(g_k)$ is a matrix derived from the shape functions at point $g_k$. To construct the overall stiffness matrix $\mathbf{K}$ for the entire tetrahedral mesh, we add each element in $\mathbf{K}_e$ computed for each tetrahedron to the corresponding entries of the overall mesh's stiffness matrix $\mathbf{K}$. This assembles a $3m \times 3m$ stiffness matrix $\mathbf{K}$.

We employ PyTorch~\citep{pytorch} to efficiently batch calculate both the element mass matrix and element stiffness matrix. Subsequently, these element matrices are assembled into global Coordinate Format (COO) sparse matrices for further processing. Notably, it is essential to highlight that these computations are automatically differentiable, enabled by PyTorch. Additionally, both the mass and stiffness matrices exhibit differentiability with respect to the material properties ($\rho$ in the mass matrix and $\mathbf{B}(E, \nu)$ in the stiffness matrix), as well as the geometry derived from our differentiable tetrahedral mesh ($N_i(x)$ in the mass matrix and $\mathbf{D}(g_k)$ in the stiffness matrix, as well as $V$ in both cases).

\subsubsection{Eigenvalue Decomposition}
\label{sec:eigenvalue_decomp}
Now, we perform a generalized eigenvalue decomposition on the mass and stiffness matrices as \(\mathbf{K} \mathbf{U} = \mathbf{M} \mathbf{U} \boldsymbol{\Lambda}\), where $\mathbf{U}$ is a stack of $k$ eigenvectors, and $\boldsymbol{\Lambda}$ is the diagonal matrix of $k$ eigenvalues. The $i$-th eigenvector, denoted as $\mathbf{u}_i$, represents the surface vibration distribution of the $i$-th mode, while the $i$-th eigenvalue, $\lambda_i$, determines its frequency and satisfies $\mathbf{K}\mathbf{u}_i = \lambda_i \mathbf{M} \mathbf{u}_i$. Since there is currently no eigenvalue decomposer that supports automatic differentiation, we thus derive the gradient relationship between eigenvalues and the mass and stiffness matrices as follows. Taking the derivative of both sides with respect to $\lambda_i$ in the equation $\mathbf{K}\mathbf{u}_i = \lambda_i \mathbf{M} \mathbf{u}_i$, we obtain:
\begin{equation}
\partial \mathbf{K} \mathbf{u}_i + \mathbf{K} \partial \mathbf{u}_i = \lambda_i \mathbf{M} \partial \mathbf{u}_i + \lambda_i \partial \mathbf{M} \mathbf{u}_i + \partial \lambda_i \mathbf{M} \mathbf{u}_i\ ,
\end{equation}
By pre-multiplying both sides by $\mathbf{u}_{i}^{T}$ and rearranging the terms, we obtain:
\begin{equation}
\mathbf{u}_{i}^{T}(\partial \mathbf{K} - \lambda_i \partial \mathbf{M}){\mathbf{u}_i} + \mathbf{u}_{i}^{T}(\mathbf{K} - \lambda_i \mathbf{M}) \partial \mathbf{u}_i = \mathbf{u}_{i}^{T} \partial \lambda_i \mathbf{M} \mathbf{u}_i.
\end{equation}
From the definition of the generalized eigenvalue, we know that $\mathbf{u}_i^T\mathbf{M}\mathbf{u}_i = 1$ and $(\mathbf{K}  - \lambda_i \mathbf{M} )\mathbf{u}_{i} = 0$. Zero coefficient ($\mathbf{u}_{i}^{T}(\mathbf{K} - \lambda_i \mathbf{M}) = 0$) automatically nullifies the gradient of the eigenvector. As a result, we can rearrange the equation and obtain:
\begin{equation}
    \partial \lambda_i = \mathbf{u}_i^T \left(\partial \mathbf{K} - \lambda_i \partial \mathbf{M}\right) \mathbf{u}_i\ .
\end{equation}

Now, we establish a connection between the gradient of vibration frequencies and the gradient of the mass and stiffness matrices. 

\vspace{0.1in}
\subsection{Loss Function for Optimization}
\label{sec:loss_func}
At this stage, we can optimize the material properties and geometry of the object using target eigenvalues. This optimization is performed by employing the loss function defined as:
\begin{equation}
L_i = ||\lambda_i^{pred} - \lambda_i^{gt} ||_1 \ ,
\end{equation}
where $\lambda_i^{gt}$ is the ground truth eigenvalue of mode $i$ and $\lambda_i^{pred}$ denotes the predicted eigenvalue. Note that the absence of a loss function regarding eigenvectors does not imply that the eigenvectors will not change. In fact, during the optimization process, gradients are transferred to the matrices $\mathbf{K}$ and 
$\mathbf{M}$. As these matrices undergo updates through gradient descent, the eigenvectors associated with them naturally evolve as well. This indicates a dynamic change in the eigenvectors even in the absence of direct modifications through the loss function.

For generality, we proceed to compute the predicted sound signal from the predicted eigenvalues as detailed in Sec. \ref{sec:additive_synthesizer}. Subsequently, we utilize a hybrid loss function to calculate the loss of the sound signal as detailed in Sec. \ref{sec:loss_func_hybrid}.

\subsubsection{Differentiable Additive Synthesizer}
\label{sec:additive_synthesizer}

The sound produced by a rigid-body object can be effectively modeled as a bank of damping sinusoidal oscillators. For the $i$-th mode, denoting its damping factor as $d_i$ and its amplitude as $A_i$, its frequency can be obtained by:
\begin{equation}
    f_i = \frac{\sqrt{\lambda_i - d_i^2}}{2 \pi} \ .
\label{eq:mode_freq}
\end{equation}
Let $h$ be the time step size, the sound signal $s_i(n)$ over discrete time steps, $n$, can be computed as:
\begin{equation}
s_i(n) = A_i e^{-d_i n h} \sin(2\pi f_i n h)\ .
\label{eq:sinusosidal_signal}
\end{equation}
Finally, the sound is produced by summing the sound signals for all modes. It is important to note that amplitudes and damping factors are designed to be learned from ground truth data, and amplitudes can implicitly include the acoustic transfer function~\citep{soundcourse}. Additionally, the eigenvalues $\lambda_i$ play a crucial role in connecting the sound signal to the physical properties of the object. The computations defined in Equations \ref{eq:mode_freq} and \ref{eq:sinusosidal_signal} are evaluated in parallel along both the time and mode dimensions using PyTorch, enabling automatic differentiation.

When dealing with naturally recorded sounds that contain noise, we enhance the output of the additive synthesizer by combining it with noise filtered by an LTV-FIR filter~\citep{engel2020ddsp}. The parameters of this filter are also learnable, enabling it to adapt to real-world noise characteristics.

\subsubsection{Hybrid Loss Function}
\label{sec:loss_func_hybrid}

As suggested in previous differential audio synthesizers~\citep{engel2020ddsp,clarke2021diffimpact}, a multi-scale spectral loss is effective for measuring the difference between two audio signals. Given the ground truth and the predicted sound signals, we compute their spectrograms $S_i$ and $\hat{S}_i$, respectively, using a specified FFT size $i$. The loss is then defined as the sum of the L1 difference between $S_i$ and $\hat{S}_i$, as well as the L1 difference between their respective log spectrograms:
\begin{equation}
L_i = ||S_i - \hat{S}_i||_1 + ||\log S_i - \log \hat{S}_i||_1 \ .
\end{equation}
The total reconstruction loss is the sum of all the spectral losses with different FFT sizes, which provide varying frequency and temporal resolutions.

Traditional L1 or L2 loss can result in difficult convergence when the initial and ground truth object locations or frequencies significantly differ \citep{xing2022}. This issue also arises in differentiable sound rendering. For instance, if the initial frequency far deviates from the ground truth frequency, there may be no overlapping pixels in the spectrogram between the initial mode and the target mode, causing the L1 or L2 loss to yield zero gradients and potentially leading to undesired local minima. 

To address this issue, we first treat the spectrogram value in each frequency bin as a high-dimensional point. To measure the distance between the ground truth and the predicted spectrograms, we utilize the optimal transport (Wasserstein) distance. This distance metric considers the cost of moving mass from one distribution to another. In our context, we define the unit moving cost from one frequency bin to another as their corresponding point distance. For efficiency, we employ an efficient algorithm for approximating optimal transport distances using Sinkhorn divergences~\citep{feydy2019interpolating}. 

As the optimal transport-based loss tends to be less effective when the initial and target spectrograms are already well-aligned, we thus propose to use a hybrid strategy: We train the model using the transport-based loss function until a plateau in loss reduction is observed, indicating no significant further decrease. At this point, we transition to employing the spectral loss function. This staged approach ensures that the model optimally benefits from each type of loss during different phases of the training process.

\section{Inverse Rendering Tasks and Experiments}

We define three types of inference tasks and conduct corresponding experiments to showcase the capability of our differentiable rendering framework. First, we perform an ablation study on the loss function to validate our approach (Sec. \ref{sec:ablation_study}). Next, we reason about the material attributes (Sec. \ref{sec:material_reasoning}), geometric shape (Sec. \ref{sec:shape_reasoning}), and impact position (Sec. \ref{sec:impact_position_reasoning}) of the object in a contact event. Please refer to the supplementary video for the demo results of our experiments.

The real-world object data used in the experiments is sourced from the \textsc{ObjectFolder Real} dataset~\citep{gao2023ObjectFolderBM}, which contains multisensory data collected from 100 real-world household objects. The data for each object includes its high-quality 3D mesh, impact sound recordings, and the accompanying video footage for each impact.
Our \name is implemented in PyTorch and utilizes the Adam optimizer for optimization.

\subsection{Ablation Study on Loss Functions}
\label{sec:ablation_study}
We first conduct an ablation study to validate the effectiveness of the hybrid loss function compared to either using a single multi-scale L1 loss or a single optimal transport-based loss. 

We set up a simple case where the predicted eigenvalues can only be changed proportionally through a trainable scaling factor. We aim to optimize this scaling factor from an initial value of 1.0 to a predefined target value. We select four meshes from the dataset and manually set the material parameters,  following the guidelines presented in~\citep{soundcourse}.

\begin{figure}[t!]
\begin{center}
    \includegraphics[width=1.0\linewidth]{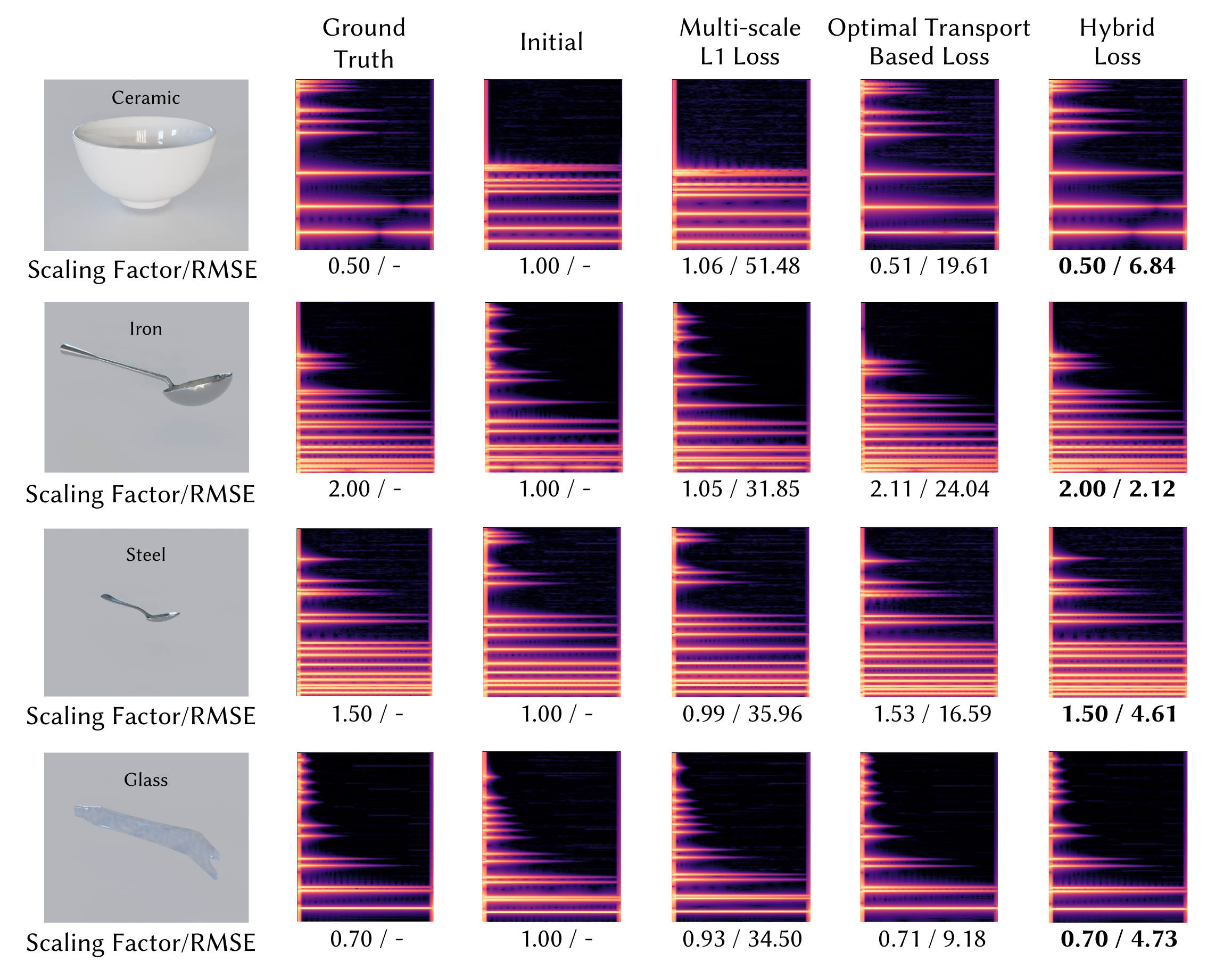}
\end{center}
  \caption{Ablation study on loss functions. We show the spectrograms, scaling factors of eigenvalues, and RMSE in different setups. Across all setups, our hybrid loss function consistently outperforms the one using only the multi-scale L1 loss or optimal transport-based loss.}
  \label{fig:loss_result}
\end{figure}

As depicted in Figure \ref{fig:loss_result}, the results indicate that the optimal transport-based loss shows high effectiveness for optimizing from a bad initial state where the multi-scale L1 loss cannot work. Additionally, our hybrid loss function achieves the best performance compared to either single loss function in all experiments.


\subsection{Material Attribute Inference}
\label{sec:material_reasoning}

In this task, we aim to infer the material parameters from the impact sound of an object, assuming the object's geometric model is known. 

\begin{figure}[h]
    \begin{center}
    \includegraphics[width=1.0\linewidth]{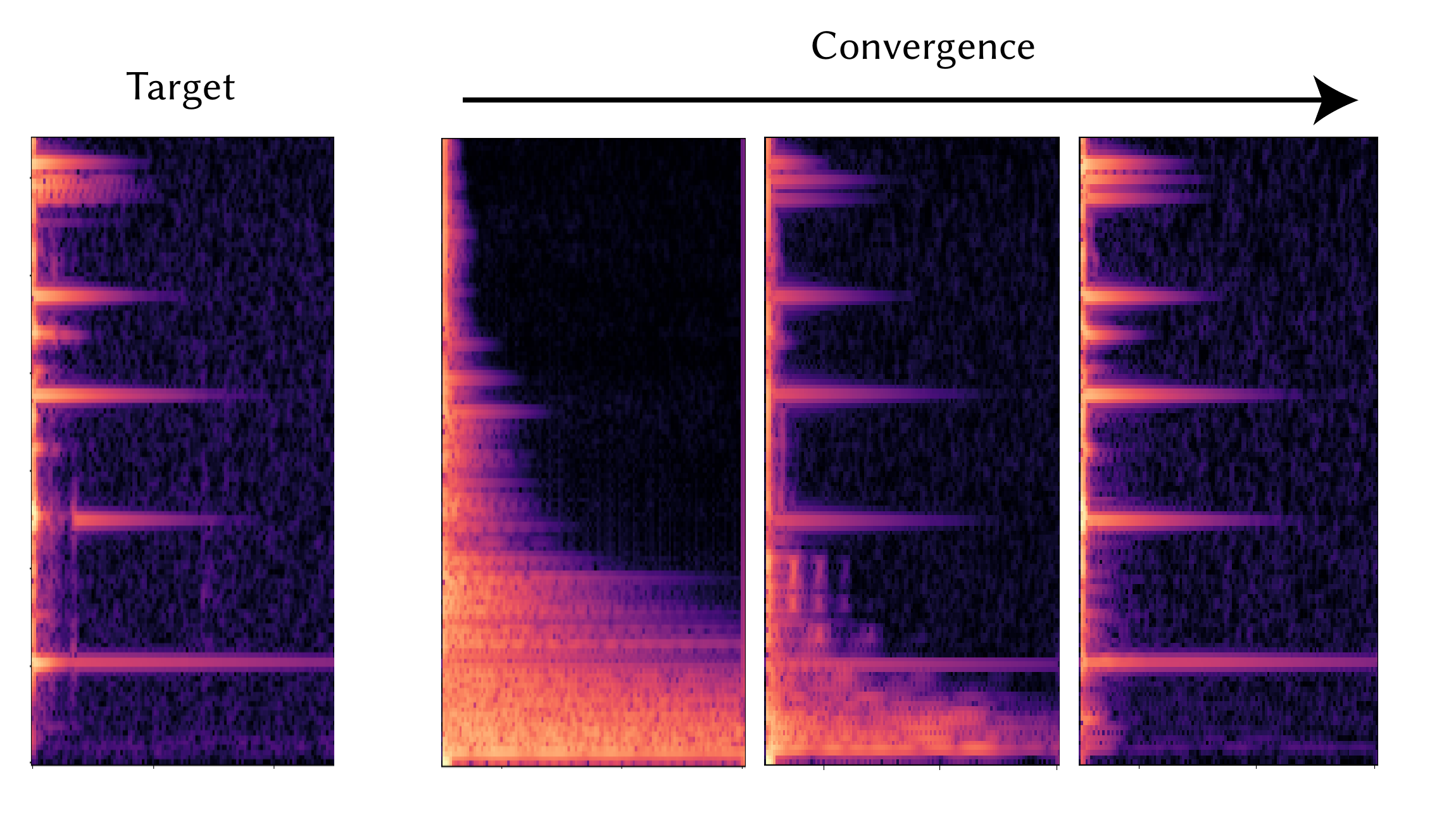}
    \end{center}
    \caption{Training process of estimating the damping curve. We utilize 256 initial modes to comprehensively cover all target modes. After training, degraded modes are subsequently removed.}
    \label{fig:damp_pretrain}
\end{figure}

First, we focus on estimating the damping curve, which is a crucial part of our differentiable additive synthesizer, denoted as \(S_{\theta_S}\) (see Figure \ref{fig:damp_pretrain}).
We define the ground truth audios as \(a_{gt}\) and use a hybrid loss function, \(L\), for optimization. For each frequency in a set of randomly selected modes, denoted as \(f_{rand}\), our synthesizer assigns a unique damping coefficient. It then uses these coefficients to synthesize the sound for these modes. The optimization of the damping curve can be formulated as:
\begin{equation}
\theta_S^* = \text{argmin}_{\theta_S}(L(S_{\theta_S}(f_{rand}), a_{gt})) \ .
\end{equation}
After optimization, we remove modes with small amplitudes and interpolate the damping coefficients of the remaining modes to form a continuous damping curve.

\begin{figure}[h]
    \begin{center}
    \includegraphics[width=1.0\linewidth]{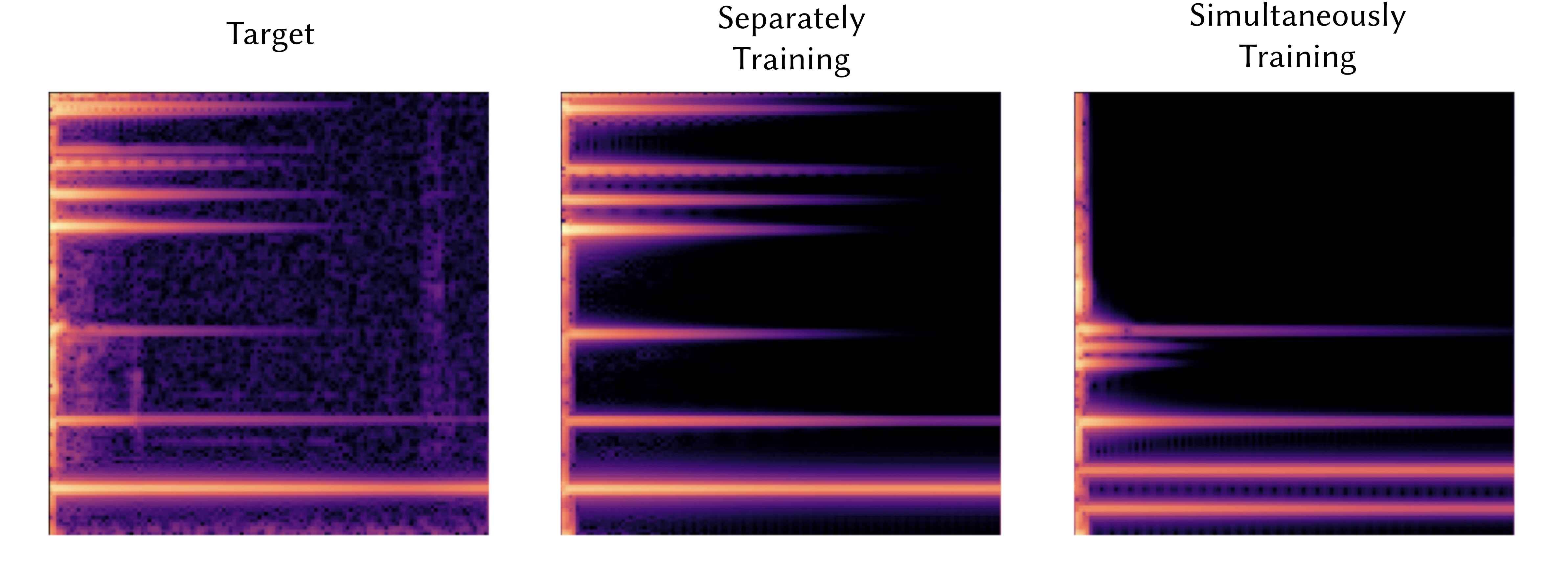}
    \end{center}
    \caption{The comparison of outcomes from separately training the damping factor and material parameters versus their simultaneous training for a test object. The images clearly demonstrate that the separate training strategy yields significantly better results than the simultaneous approach, demonstrating the effectiveness of our separate training strategy.}
    \label{fig:damping_curve_compare}
\end{figure}

Next, we estimate the material parameters of our FEM matrix assembler, \(A_{\theta_A}\). These parameters include the ratio of Young's modulus to density (denoted as \(\hat{E}\)) and Poisson's ratio (denoted as \(\nu\)). Our framework \(F_\theta\) integrates several components: a fixed tetrahedron mesh generator \(G\), the FEM matrix assembler \(A_{\theta_A}\), an eigenvalue decomposer \(D\), and the previously trained and now fixed additive synthesizer \(S_{\theta_S^*}\). For a given input model \(m\), the optimization of the material parameters can be expressed as:
\begin{equation}
\theta_A^* = \text{argmin}_{\theta_A}(L(F_\theta(m), a_{gt}))\ .
\end{equation} We separate the training processes for the damping factor and material parameters. This is because the errors in the damping factor and material parameters can influence each other, making optimization challenging. In this two-step process, we sequentially refine the estimation of the damping and material parameters, leading to a more accurate and stable optimization. Figure \ref{fig:damping_curve_compare} presents a comparison of the  the outcomes from training the damping factor and material parameters separately versus training them simultaneously on a test object. The results demonstrate that our approach of separate training substantially outperforms the simultaneous training method, proving to be highly effective.


Contrasting with prior work \citep{Ren2013} that used first-order FEM with a fixed Poisson's ratio to predict only Young's modulus, their model's eigenvalues are simplified with a linear proportional relationship with Young's modulus. In contrast, our method optimizes both Young's modulus and Poisson's ratio, addressing the inaccuracies arising from the oversimplified assumption. We provide a detailed comparison with various baselines in Table \ref{tab:material_estimate_result}.

We leverage ground-truth audio data, synthesized using second-order FEM for 16 objects with randomly chosen material parameters from a feasible range. Beginning with random initial material parameters, our framework differentiably synthesizes sound by optimizing these parameters to minimize the loss against the ground truth. Moreover, the effectiveness of our approach is evaluated using data from two real-world ceramic objects.

\begin{figure*}[t]
    \centering
    \includegraphics[width=\linewidth]{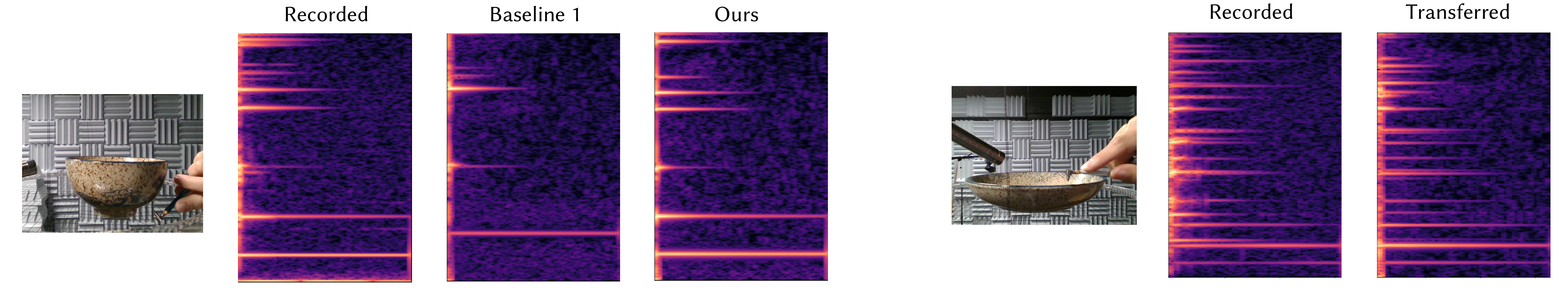} 
    \vspace{-0.1in}
    \caption{(Left) Material estimation from real-world recorded sound with our \name method and the baseline of \citep{Ren2013}, which uses a fixed value of Poisson's ratio in first-order FEM. (Right) Transfer of the material parameters optimized from a ceramic bowl to a plate with the same material, with additional fine-tuning of the noise filter and mode amplitude.}
    \vspace{-0.1in}
    \label{fig:transfer_result}
\end{figure*}

We use the relative error as a metric for $\hat{E}$, $\nu$, and sound spectrogram, defined as $l = \frac{||g - p||_2}{||g||_2}$ for the ground-truth $g$ and the prediction $p$. We present the quantitative results in Table \ref{tab:material_estimate_result} for synthetic data, along with qualitative examples for real-world data in Figure \ref{fig:transfer_result}.  Our \name demonstrates substantial improvements over all baselines across all metrics, showcasing high effectiveness even in real-world data. 

\paragraph{Implementation details}
To estimate the damping curve, we initially train an additive synthesizer using target audios composed of 256 modes with filtered noise. The training spans over 10,000 steps in few minutes. Modes with damping coefficients less than 100 are deemed invalid, and the coefficients of the remaining valid modes are linearly interpolated to form a damping curve. Subsequently, we train the material parameters based on the damping curve, spanning 10,000 steps at a learning rate of 0.01, with a transition in the loss function at the 5,000-step mark. Using a background tetrahedral mesh grid resolution of $32^3$, the total training duration for each object is approximately 3 hours.

\begin{table}
    \begin{center}
    \caption{Material estimation with 16 objects. Baseline 1 \cite{Ren2013} uses fixed Poisson's ratio and first-order FEM; Baseline 2 applies 2nd-order FEM with fixed Poisson's ratio; Baseline 3 employs 1st-order FEM with learnable Poisson's ratio. Our method consistently outperforms all baselines in relative errors.
}
\scalebox{0.9}{
\begin{tabular}{lccccc}

\toprule
                     & FEM order & Learnable $\nu$ & $\hat{E}$ Err. & $\nu$ Err. & Spec. Err. \\ 
\midrule
baseline 1           & 1         &    \ding{55}        &        0.51        &      0.68      &      26.43      \\
baseline 2           & 2         &     \ding{55}         &      0.10         &    0.68        &    11.21        \\
baseline 3           & 1         &     \checkmark        &      0.51          &     0.66       &    27.00        \\
\name & 2         &    \checkmark      &    \textbf{0.07}       &     \textbf{0.26}       &      \textbf{7.95}      \\ 
\bottomrule
\end{tabular}
}
    \label{tab:material_estimate_result}
    \end{center}
\end{table}

\subsection{Shape Attribute Inference}
\label{sec:shape_reasoning}
Determining the shape from sound is challenging because different shapes can produce similar sounds upon impact~\citep{kac1966can}. Therefore, we regulate the material coefficients and impose certain geometry constraints to ensure a reliable optimization process.

\subsubsection{Geometric Shape Inference}


In this task, our focus is on detailed geometry recovery under certain constraints. Our goal is to precisely reconstruct the finer details of shapes from a given coarse voxel grid by utilizing modal sound for inference.

To accomplish this task, we infer the geometric shape from the eigenvalues of vibration modes, which are directly related to frequencies (\autoref{eq:mode_freq}). Additionally, we constrain the tetrahedral mesh during optimization using a coarse voxel grid. Specifically, we query the SDF values from the MLP and ensure that the SDF of grid points inside the mesh is negative, while those outside are positive. This is enforced using a loss defined as the sum of absolute SDF values of those points whose SDF sign differs from the expected sign. The loss for sound constraint is defined as the L1 loss between the ground truth eigenvalues and the predicted eigenvalues of the first $k$ modes, divided by the norm of the ground truth. 

In our framework, the key learnable parameter, denoted as \(\theta\), is the implicit SDF representation, \(SDF_\theta\), in the tetrahedron mesh generator \(G_{\theta_G}\). This generator processes the coordinates of a point to output its SDF value for further synthesis. Our framework integrates the tetrahedron mesh generator \(G_{\theta_G}\), the fixed FEM matrix assembler \(A\), and the eigenvalue decomposer \(D\), yielding the function \(F_\theta\). This function takes an initial coarse mesh \(m\) as input and outputs the first \(k\) smallest eigenvalues associated with it.

Let \(\lambda_{gt}\) represent the ground truth eigenvalues and \(SDF\) denote the SDF field defined by the initial coarse voxels. The weight of the coarse voxel grid constraint is given by \(w\). For all vertices \(v\) in the background grid \(B\), the optimization process can be formulated to minimize the combination of the eigenvalue loss and the SDF sign discrepancy loss:
\begin{equation}
\theta^* = \text{argmin}_\theta (L_{\text{eigen}}(\theta) + w \cdot L_{\text{SDF}}(\theta)) \ .
\end{equation}
The eigenvalue loss \(L_{\text{eigen}}(\theta)\) is defined as:
\begin{equation}
L_{\text{eigen}}(\theta) = \lVert F_\theta(m) - \lambda_{gt} \rVert_1 \ .
\label{eq:geo_optimize}
\end{equation}
The SDF sign discrepancy loss \(L_{\text{SDF}}(\theta)\) is defined over the background grid \(B\) to penalize the discrepancy in the signs of the SDF values between the predicted and initial coarse voxel fields as:
\begin{equation}
L_{\text{SDF}}(\theta) = \sum_{v \in B} |SDF_\theta(v)| \cdot | \text{sign}(SDF_\theta(v)) - \text{sign}(SDF(v)) |
\end{equation}
In this formulation, \(\theta^*\) is the optimized parameter set that minimizes the combined loss.

\begin{figure*}[h]
    \centering
    \includegraphics[width=1.0\linewidth]{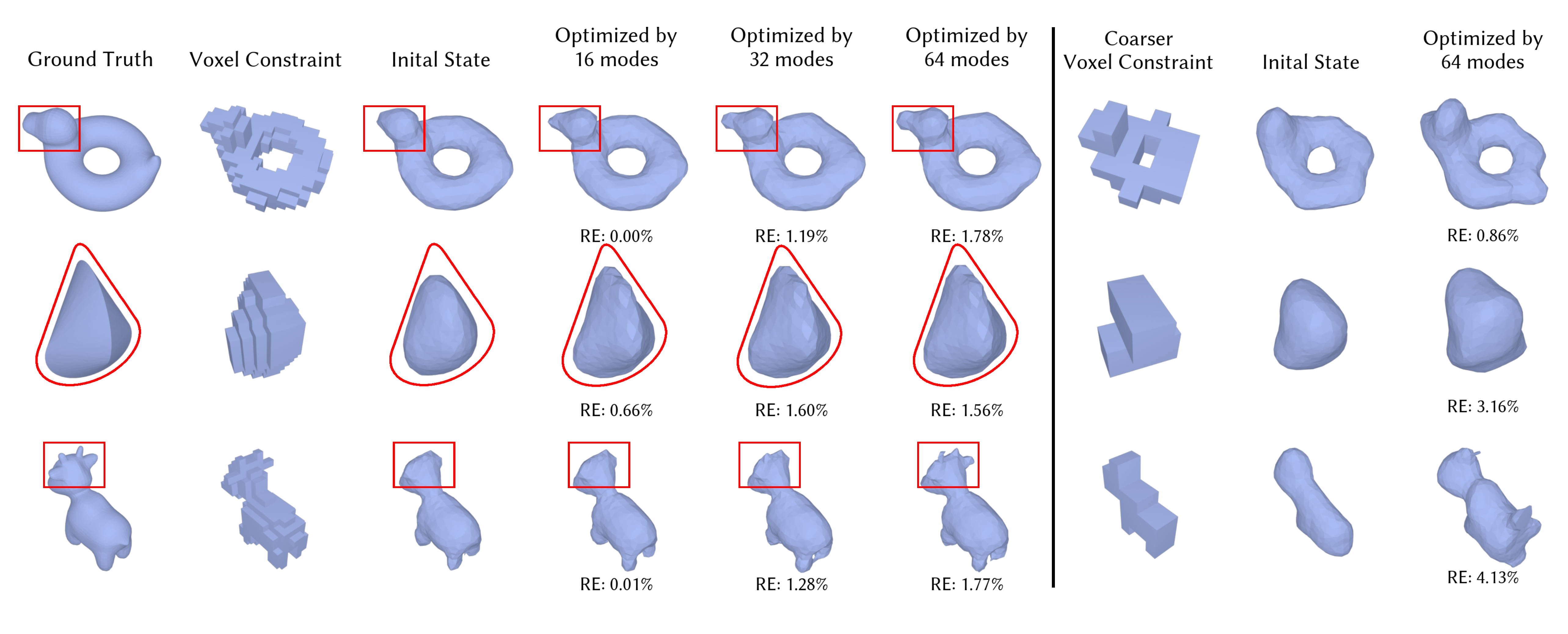} 
    \vspace{-0.2in}
    \caption{Optimizing shape detail through sound mode (eigenvalues of the ground-truth model) and voxel grid ($16^3$). Our approach demonstrates its capability to restore shape details with small Relative Error (RE) of eigenvalues. With an increase in the number of modes, fitting all modes simultaneously becomes challenging, causing an increase in RE. However, more modes enforce stricter constraints on shape optimization, yielding an optimized mesh that more closely resembles the ground truth in detail. Three rightmost columns highlight our outcomes when applying coarser voxel constraints ($8^3$). Our approach can visually synthesize geometric details that closely resemble the ground truth.}
    \label{fig:shape_result}
\end{figure*}

In our experiments, we generate synthetic data for three objects from~\citet{crane2013robust} with a ceramic material parameter. We conduct separate experiments for each object and constraint mode number. The geometric shape can be successfully recovered from impact sound and a coarse voxel constraint ($16^3$), as illustrated by the quantitative results in Figure \ref{fig:shape_result}. This capability compensates for the loss of such details in the initial coarse mesh. When applied to coarser voxel constraints ($8^3$), our approach can still synthesize geometric details that are visually close to the ground truth. The high quality of shape estimation of our approach can also be validated in the accompanying video.

\paragraph{Implementation details}
First, we obtain the initial MLP representation for the SDF by optimizing with only the constraint of a coarse voxel grid. The SDF constraint involves $32^3$ uniformly distributed points within a cube that is 1.1 times the size of the object's bounding box. This initial training process requires only a few seconds for 1000 steps. Subsequently, we refine the MLP representation by training according to \autoref{eq:geo_optimize} with $\omega = 0.001$. Using a background tetrahedral mesh grid resolution of $32^3$, this process spans 100 steps and takes approximately 2 minutes.

\subsubsection{Volumetric Thickness Inference}

An interesting question arises: Is a given object solid or hollow? And if so, how thick is the inside? In this task, our focus shifts to understanding the internal conditions of an object as a whole. Our goal is to precisely estimate the volumetric thickness of a given object by utilizing modal sound for inference.

To complete this task, we generate hollow meshes from each solid input, varying only in thickness, and then predict their thickness based on their eigenvalues.
Mesh "thickness" is defined using the solid mesh's Signed Distance Field (SDF): the smallest SDF value, \(s_{\min}\), corresponds to the farthest internal point from the outer surface. An object of thickness \(t\) includes points within \(t \cdot -(s_{\min})\) from this surface. Thus, a point \(P\) lies inside an object of thickness \(t\) if its SDF value \(SDF(P)\) satisfies \(-t \cdot s_{\min} < SDF(P) < 0\).



Based on the defined "thickness", we enhanced our tetrahedral mesh generator \( G_{\theta_G} = G_t \) with a thickness parameter \( t \), enabling it to synthesize hollow voxel meshes of thickness \( t \) from a solid mesh. We chose several target thicknesses \( t_{target} \) for each input mesh and created corresponding ground truth meshes for each. During optimization, \( G_t \) differentiably generates a hollow voxel mesh from the initial SDF of the mesh, matching the current thickness \( t \). We then compute the L2 loss between the first \( k \) eigenvalues of the current and ground truth meshes, updating \( t \) based on its gradient.

We consolidate the tetrahedron mesh generator \(G_t\), the fixed FEM matrix assembler \(A\) and the eigenvalue decomposer \(D\) in our framework into a function \(F_t\), which takes an input mesh and outputs first \(k\) eigenvalues of its voxel corresponding tetrahedron mesh with thickness \(t\). Assuming ground truth eigenvalues is \(\lambda_{gt}\), for initial solid mesh \(m\), formulation of this optimization process can be expressed as:
\begin{equation}
t^* = \text{argmin}_t(||F_t(m) - \lambda_{gt}||^2) \ .
\end{equation}

In our experiment, we selected four models with varying materials. We generated target grids for each model at thicknesses of 0.3, 0.4, 0.5, 0.6, and 0.7, and used their first \(k=32\) eigenvalues as ground truth. We observed that thicknesses below 0.3 led to holes in the mesh due to limited resolution, while thicknesses above 0.7 yielded eigenvalues akin to those of solid objects. The results, detailed in Table \ref{tab:thickness_estimate_result}, indicate our method's proficiency with smooth, thick meshes, but potential inaccuracies with complex surfaces or thinner meshes. This can be attributed to a more complex nonlinear relationship between eigenvalues and thickness in such cases.

\paragraph{Implementation details}
Using a background tetrahedral mesh grid resolution of $64^3$, the training process encompasses 500 steps at a learning rate of 0.02, taking approximately 2 hours per object and target thickness.

 \begin{table}
    \begin{center}
        \caption{Volumetric thickness inference using synthetic data tested with different objects. We measure the prediction error against the ground-truth coefficient using MAE. Our approach achieves high prediction accuracy.}
\begin{tabular}{lccccccc}
\toprule

\multicolumn{1}{l}{\multirow{2}*{object}} & \multicolumn{5}{c}{target thickness}  & \multicolumn{1}{l}{\multirow{2}*{MAE}} &                                    \\ \cline{2-6}
\multicolumn{1}{l}{} & 0.3 & 0.4 & 0.5 & 0.6 & 0.7 & \multicolumn{1}{l}{} \\

\midrule
Bunny &  0.304         &    0.407        &        0.508      &      0.608      &      0.709  & 0.0073    \\
Armadillo & 0.338         &     0.456         &      0.590        &    0.696        &    0.730  & 0.0623      \\
Bulbasaur & 0.308         &     0.411        &      0.512         &    0.614       &    0.718 & 0.0125       \\
Squirtle &  0.312      &    0.416       &     0.520       &      0.624 & 0.718 & 0.0177 \\ 
\bottomrule
\end{tabular}
    \label{tab:thickness_estimate_result}
    \end{center}
\end{table}

\subsubsection{Shape Morphing Inference}
\begin{figure*}[h]
    \centering
\includegraphics[width=1.0\linewidth]{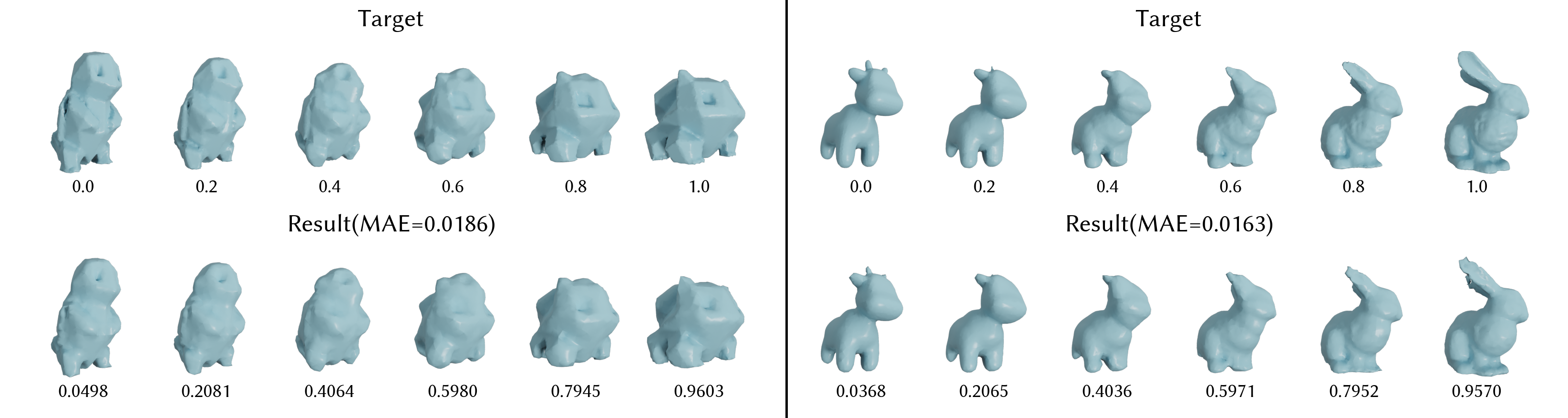}
    \caption{Shape morphing recovery from successive meshes. We annotate each shape with its corresponding target/predicted interpolation coefficients. The prediction accuracy of these coefficients is assessed using Mean Absolute Error (MAE). Our results demonstrate that our framework effectively predicts the mesh nearest to the target eigenvalue, along with its associated shape.}
\vspace{-0.1in}
\label{fig:interpolate_result}
\end{figure*}

In this task, we focus on two distinct objects and a series of intermediate morphing shapes between them. Our objective is to accurately identify the specific shapes within this progression, utilizing modal sound for inference.
To accomplish this task, we begin by calculating the SDF values for each vertex in the DMTet background grid of the two initial meshes. We then interpolate the SDF values from these meshes for each vertex. Let \(v_{SDF1}\) and \(v_{SDF2}\) be the SDF values for a background grid vertex \(v\) relative to the two models, and \(t\) be the interpolation coefficient. The vertex's interpolated SDF value is \(v_{SDF} = t \cdot v_{SDF1} + (1 - t) \cdot v_{SDF2}\). As \(t\) varies from 0 to 1, the mesh transitions from the first to the second mesh, as depicted in the Target part in \autoref{fig:interpolate_result}. Our objective is to find the mesh that optimally fits the target sound during this morphing, by optimizing the interpolation coefficient \(t\).


For two initial meshes \(m_1\) and \(m_2\),   \(SDF(\cdot)\) gives its SDF values in all background grid vertexes \(SDF(m_1)\) and \(SDF(m_2)\). We consolidate the tetrahedron mesh generator \(G_{\theta_G} = G_t\), the fixed FEM matrix assembler \(A\), and the eigenvalue decomposer \(D\) in our framework into a function \(F_t\), which takes background grid SDF and outputs its corresponding first \(k\) eigenvalues.

Assuming ground truth first \(k\) eigenvalues is \(\lambda_{gt}\), The formulation of this optimization process can be expressed as:
\begin{equation}
t^* = \text{argmin}_t(||F_t(t \cdot SDF(m_1) + (1-t) \cdot SDF(m_2)) - \lambda_{gt}||^2).
\end{equation}

To validate the feasibility of our framework for this task, we select a series of initial mesh pairs, select a range of target interpolation coefficients, interpolate SDF values from target coefficients, synthesize morphed tetrahedral voxel meshes from SDF values, and extract its first \(k\) eigenvalues as ground truth. During the optimization process, we start from randomly initialized interpolation coefficients \(t\), synthesize the interpolated meshes from current interpolation coefficients, calculate the L2 loss between their eigenvalues and the ground truth, and backpropagate gradients to update the interpolation coefficients \(t\).

In our experiment, we used Squirtle and Bulbasaur, and Bunny and Spot, as two pairs of initial meshes for morphing. We interpolated models using SDF coefficients of 0.0, 0.2, 0.4, 0.6, 0.8, and 1.0, and extracted their first \(k=32\) eigenvalues as shape morphing targets. To keep the interpolation coefficient \(t\) between 0 and 1, \(t\) was represented as a weighted sum of 16 evenly spaced numbers from 0 to 1. These weights, treated as learnable parameters, were normalized to be positive and sum to 1. Figure \ref{fig:interpolate_result} visualizes the results, showing our framework's efficiency in finding shapes that align closely with the target eigenvalues during transformation. The implementation details are exactly the same as those for the experiments on thickness inference except for the learning rate, which is set to be 0.05.


\subsection{Impact Event Inference}
\label{sec:impact_position_reasoning}
Impact position and amplitude are not explicitly optimized as a learnable parameter. However, the learnable mode amplitude \(A\) in Equation~\ref{eq:sinusosidal_signal} implicitly encodes information about the impact position and amplitude. In other words, the mode amplitude can be utilized to predict the object's impact position and amplitude, a conclusion that has also been reached in ~\citet{van1998sounds}.

\begin{figure}[b!]
    \centering
\includegraphics[width=0.8\linewidth]{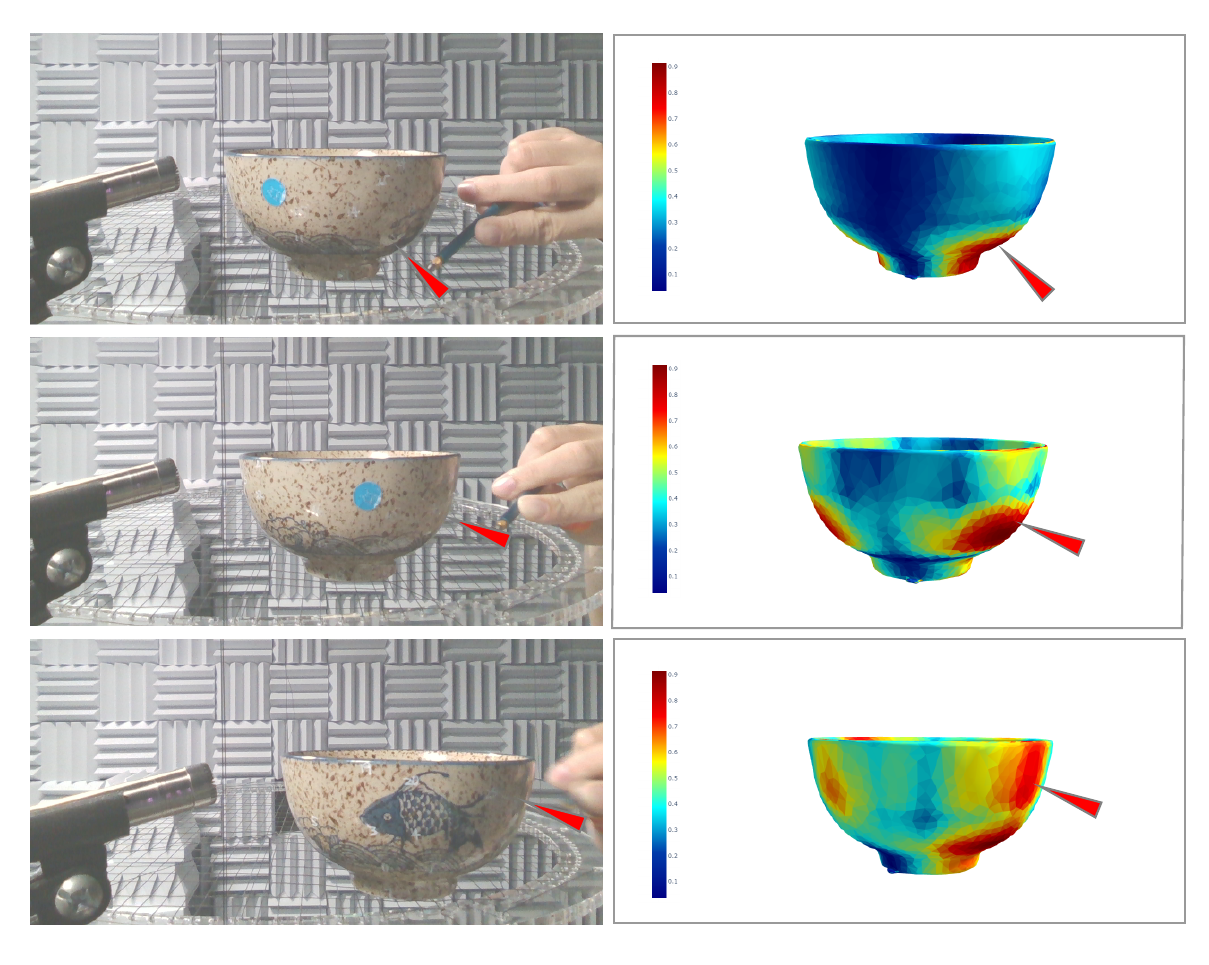}
    \caption{Visualization of the surface likelihood distribution (probability heatmap) of the impact position on the object's surface for an example object. The predicted positions are considered reasonable and accurate if they fall within the region that is rotationally symmetric about the central axis relative to the groundtruth. }

\label{fig:impact_position}
\end{figure}

In this task, we aim to infer the impact position from the recorded sound, given that the object's mesh is known. First, we optimize the material parameters from sound following the process outlined in Sec. \ref{sec:material_reasoning}. Simultaneously, we optimize the amplitudes of all modes, denoted as $\mathbf{A} = [A_0, A_1, ..., A_n]$. Then, using the estimated material parameters, we apply forward modal sound simulation, which includes acoustic transfer~\citep{james2006precomputed}, to obtain the simulated amplitudes of all modes $\hat{\mathbf{A}}_i$ when impacting each mesh vertex $v_i$. We measure the likelihood that the impact position corresponding to the recorded sound is near vertex \( v_i \) by evaluating the similarity between \(\mathbf{A}\) and \(\hat{\mathbf{A}}_i\). We choose recorded real data of a ceramic bowl from the \textsc{ObjectFolder Real} dataset~\citep{gao2022ObjectFolderV2} and use cosine similarity to compute the surface likelihood distribution, as visualized in Figure~\ref{fig:impact_position}. In this example, the bowl exhibits rotational symmetry around its central axis. Theoretically, striking any points on the bowl that are rotationally symmetric about this axis (forming a circle centered on the axis) would produce identical sounds with symmetric impulse responses. The impact positions predicted by the mode amplitudes are deemed reasonable and accurate if they fall within the same circle as the groundtruth (i.e., they are rotationally symmetric about the central axis). For ease of visual comparison, we rotate the probability heatmap of the predicted impact positions to align them with the groundtruth. And our method predicts a high likelihood around the ground truth impact position.

\section{Conclusion}
We presented a differentiable modal sound rendering framework that enables inverse rendering by computing the gradient of the simulation function with respect to input physical parameters (e.g., material parameters). We have verified the effectiveness of our loss strategy with ablation experiments and demonstrated the generality and diversity of \name for the inverse rendering of material parameters, impact positions, and the shape of the sounding objects. We hope our framework can unlock new multisensory applications in the fields of robotics and embodied AI.

Nonetheless, our framework currently faces several challenges. These include difficulties in handling complex shapes, particularly thin shells, and limitations in accurately modeling heavily nonlinear sounds. Additionally, if the distribution of an object's damping coefficients fluctuates significantly from low to high frequencies, a simple interpolated damping curve may fail to accurately represent the true distribution. Furthermore, optimizing the rendering speed to support real-time applications remains a critical priority. In future endeavors, we aim to develop a more comprehensive and efficient differentiable sound rendering framework, building upon the foundation laid by this work.


\begin{acks}
We thank all the anonymous reviewers for their helpful suggestions. This work is supported by the \grantsponsor{nkpc}{National Key R\&D Program of China}{} (No.~\grantnum{nkpc}{2022YFB3303403}) and~\grantsponsor{nsfc}{NSFC of China}{} (No.~\grantnum{nsfc}{62172013}). 
\end{acks}

\bibliographystyle{ACM-Reference-Format}
\bibliography{diffsound}

\clearpage

\end{document}